\documentclass[aps,pra,reprint,superscriptaddress,showpacs]{revtex4-2}
\usepackage{graphicx}
\usepackage{dcolumn}
\usepackage{bm}
\usepackage{subfigure}
\usepackage{algorithm}
\usepackage{algpseudocode}
\usepackage{appendix}
\usepackage{float}
\usepackage[braket, qm]{qcircuit}
\usepackage{mathrsfs}
\usepackage{longtable}
\usepackage{amsfonts}
\usepackage{amstext}
\usepackage[usenames]{xcolor}
\usepackage{epsfig}
\usepackage{epstopdf}
\usepackage{mathrsfs}
\usepackage{bm}
\usepackage{amsmath}
\usepackage{amssymb,amsthm}
\usepackage{tikz}
\usepackage{hyperref}
\usepackage{balance}

\usepackage{comment}

\hypersetup{colorlinks=true, citecolor=blue, urlcolor=blue, linkcolor=blue}

\newtheorem{theorem}{Theorem}

\newtheorem*{pf}{Proof}
\newtheorem{df}{Definition}

\begin{document}

\title{Static and dynamic coherence fraction in the Bernstein-Vazirani algorithm}
\author{Si-Qi Zhou}
\affiliation{School of Mathematical Sciences, MOE-LSC, Shanghai Jiao Tong University, Shanghai, 200240, China}
\affiliation{Shanghai Seres Information Technology Co., Ltd, Shanghai, 200040, China} \affiliation{Shenzhen Institute for Quantum Science and Engineering, Southern University of Science and Technology, Shenzhen, 518055, China}
\author{Jin-Min Liang}
\affiliation{State Key Laboratory for Mesoscopic Physics, School of Physics, Frontiers Science Center for Nano-optoelectronics, $\&$ Collaborative Innovation Center of Quantum Matter, Peking University, Beijing 100871, China}
\author{Jiayin Peng}
\affiliation{School of Math. Inform. Sci. Neijiang Normal University, Neijiang 641100, Sichuan, China}
\author{Zhihua Chen}
\email{chenzhihua77@sina.com}
\affiliation{School of Science, Jimei University, Xiamen 361021, China}
\author{Shao-Ming Fei}
\email{smfei@mis.mpg.de}
\affiliation{School of Mathematical Sciences, Capital Normal
University, Beijing 100048, China}
\affiliation{Max Planck Institute for Mathematics in the Sciences - 04103 Leipzig, Germany}
\author{Zhihao Ma}
\email{mazhihao@sjtu.edu.cn}
\affiliation{School of Mathematical Sciences, MOE-LSC, Shanghai Jiao Tong University, Shanghai, 200240, China}
\affiliation{Shanghai Seres Information Technology Co., Ltd, Shanghai, 200040, China}
\affiliation{Shenzhen Institute for Quantum Science and Engineering, Southern University of Science and Technology, Shenzhen, 518055, China}

\begin{abstract}
Quantum entanglement and coherence are crucial resources in quantum information theory. In some scenarios, however, it is not necessary to directly estimate entanglement or coherence measures to quantify the capabilities of a state in quantum information processing. Instead, fully entangled fraction and coherence fraction are two alternatives for entanglement and coherence in specific quantum tasks. Here, we establish a link between the coherence fraction and the Bernstein-Vazirani algorithm, which has several potential applications including cryptography and database search. We show that the success probability of the generalized Bernstein-Vazirani algorithm depends only on the coherence fraction of the initial state rather than its entanglement or coherence. Moreover, we discuss the coherence fraction dynamics and establish a relation between the operator's coherence fraction and the algorithm's success probability. Our findings highlight how quantum coherence fraction influences the efficiency of quantum algorithms.
\end{abstract}
\maketitle
\section{Introduction}
Quantum entanglement~\cite{PhysRevLett.78.2275,RevModPhys.81.865} is one of the prominent resources in quantum theory, enabling tasks that are either superior or impossible by classical means~\cite{PhysRevLett.69.2881,PhysRevLett.87.077902,PhysRevLett.87.047901,PhysRevLett.110.060504,PhysRevX.5.041008}. At the core of entanglement lies the coherent superposition of states, which can be viewed as a specific manifestation of quantum coherence~\cite{giovannetti2004quantum,giovannetti2011advances,PhysRevX.5.021001,lostaglio2015description}. The resource theory of entanglement has a long tradition. In contrast, the coherence theory was formalized more recently~\cite{PhysRevLett.113.140401}. Quantum coherence is recognized as another key resource in quantum information processing, playing a crucial role in the advancement of quantum physics~\cite{PhysRevLett.115.020403,du2015coherencemeasuresoptimalconversion,PhysRevLett.116.080402,PhysRevLett.116.120404,PhysRevLett.116.160407,PhysRevLett.116.150502,PhysRevLett.117.030401,RevModPhys.89.041003,PhysRevLett.120.230504,hu2018quantum,wu2021experimental,PhysRevLett.132.180202,PhysRevA.109.052443,PhysRevA.110.042425,zhang2024quantification}. Both quantum entanglement and coherence have been demonstrated to play crucial roles in enhancing the performance and efficiency of quantum algorithms~\cite{jozsa1998quantum,ekert1998quantum,PhysRevA.65.062312,PhysRevA.93.012111,PhysRevA.95.032307,PhysRevA.100.012349,pan2019entangling,pan2022complementarity,PhysRevA.106.062429,PhysRevLett.129.120501}.

While entanglement and coherence are essential resources in quantum information theory and quantum algorithms, direct evaluation of entanglement or coherence measures is not always required to assess the capabilities of a state for specific tasks. Quantifiers such as the fully entangled fraction (FEF)~\cite{PhysRevA.54.3824} and quantum coherence fraction (QCF)~\cite{PhysRevA.100.032324} can serve as effective indicators. The FEF in entanglement theory quantifies the entanglement content of a state by measuring the overlap between a state and a maximally entangled state, playing a key role in processes like dense coding~\cite{grondalski2002fully}, teleportation~\cite{PhysRevA.62.012311,PhysRevA.66.012301,PhysRevLett.127.080502}, nonlocal correlations~\cite{PhysRevA.94.062120,PhysRevA.94.062123}, and even quantum thermodynamics~\cite{PhysRevA.96.012107}. Similarly, the QCF in coherence theory quantifies the overlap of a given quantum state with a maximally coherent state~\cite{bai2015maximallycoherentstates,PhysRevA.93.032326,karmakar2019coherence}. This quantity not only establishes a direct link to operational tasks, such as quantum coherence distillation~\cite{PhysRevA.92.022124}, but also aligns closely with the framework of resource theory. Furthermore, QCF exhibits connections to other coherence measures, including the robustness of coherence~\cite{PhysRevLett.116.150502} and the $l_{1}$-norm coherence~\cite{PhysRevLett.113.140401, karmakar2019coherence}. Notably, QCF is computationally efficient for low-dimensional systems and provides a tight bound for the robustness of coherence, making it particularly valuable for practical quantum applications~\cite{PhysRevA.100.032324}. Recently, reference~\cite{PhysRevA.110.062429} demonstrated that the success probability of the Grover search algorithm is highly related to the QCF of the initial state. 

Bernstein-Vazirani (BV) algorithm~\cite{bernstein1993quantum,bernstein1997quantum} identifies an unknown bit string encoded as a linear function using only an oracle. In contrast, the classical approach requires $n$ oracle queries. By extending the Deutsch–Jozsa (DJ) algorithm~\cite{deutsch1985quantum,deutsch1992rapid}, the BV algorithm shows that quantum algorithms are capable of determining not only the properties but also the exact form of Boolean functions~\cite{xie2018quantum,zhou2023distributed}. Recently, Bravyi et al.~\cite{bravyi2018quantum} introduced a non-oracular version of the BV algorithm, demonstrating that constant-depth quantum circuits are more powerful than their classical counterparts. Naseri et al.~\cite{PhysRevA.106.062429} conducted a rigorous quantitative investigation of quantum resources in the probabilistic BV algorithm, demonstrating that, without entanglement in the initial and the final states, the performance is directly related to the amount of coherence in the initial state. More recently, Pokharel et al.~\cite{PhysRevLett.130.210602} explored a single-shot version of the BV algorithm which offers a provable, conjecture-free exponential speedup over the best possible classical algorithm. These contributions highlight the evolving understanding of quantum advantage in the BV algorithm.

\begin{figure}[H]
\centering
\includegraphics[width=2.5in]{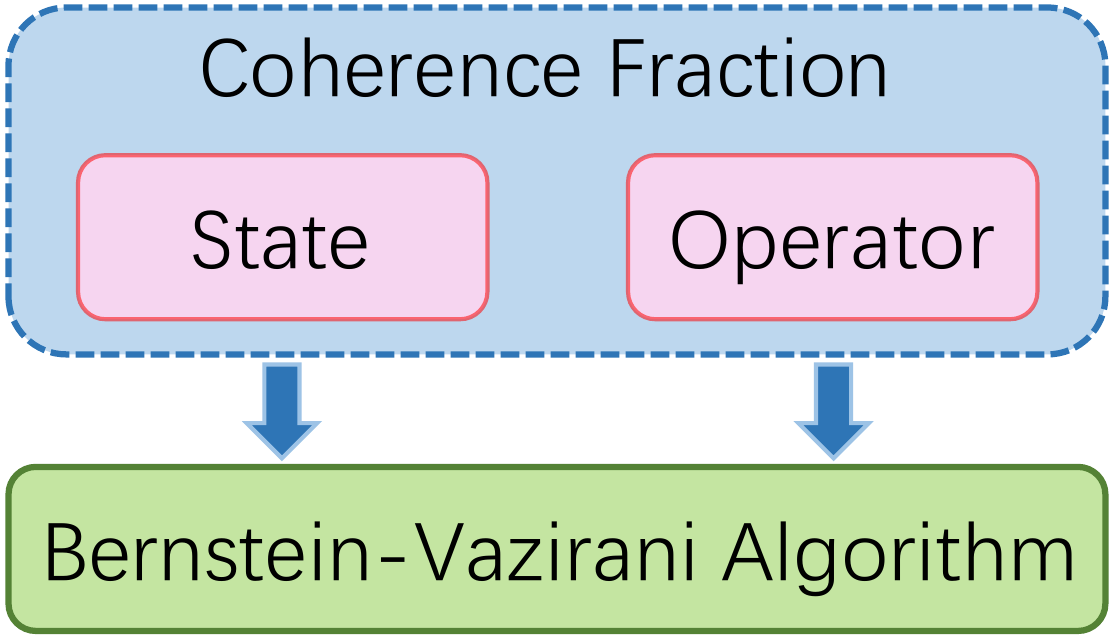}
\caption{\textbf{Diagrammatic sketch of quantum coherence fraction in the Bernstein-Vazirani algorithm.} The connections between the success probability of a generalized Bernstein-Vazirani algorithm and the coherence fraction, including both state and operator coherence fractions.}
\label{fig:CFBV}
\end{figure}

In this work, we propose a generalized version of the BV (GBV) algorithm and investigate the connections between the success probability of this algorithm and the QCF, covering both state and operator coherence fractions. A sketch diagram is provided in FIG.~\ref{fig:CFBV}. Our results reveal that the success probability of the GBV algorithm does not depend on the entanglement or coherence of the initial state. Instead, it depends on the coherence fraction of the initial state. To clarify the role of the coherence fraction at the operator level in the algorithm, we explore the dynamics of the state after each operator is applied and establish a direct link between the coherence fraction of each operator and the success probability of the GBV algorithm. We also give a detailed example to illustrate the role of the QCF in the GBV algorithm.

Our paper is organized as follows. In Sec.~\ref{II}, we recall the BV algorithm and propose the GBV algorithm which is a generalized version of the BV algorithm. Sec.~\ref{III} investigates the connections between the success probability of the GBV algorithm and the QCF, covering both state and operator coherence fractions. We give a detailed example to illustrate the QCF dynamics in Sec.~\ref{IV}. Discussions are given in Sec.~\ref{V}.
\section{Bernstein-Vazirani algorithm}\label{II}
Our starting point is the well-known Bernstein-Vazirani (BV) problem~\cite{bernstein1993quantum,bernstein1997quantum}. Here one is given an oracle access to a linear boolean function $\ell: \mathbb{F}_{2}^{n} \rightarrow \mathbb{F}_{2}$ parameterized by a “secret” bit string $z \in \{0,1\}^{n}$, such that $\ell(x)=z \cdot x \bmod 2$, $x \in\{0,1\}^{n}$. Here and below $z \cdot x \equiv \sum_{k=1}^{n} z_k x_k$ denotes the inner product of vectors~\cite{bravyi2018quantum}. 

Bernstein and Vazirani showed that one can identify the linear function $\ell$ and find the secret bit string $z$ by using just one quantum query to an oracle $\mathcal{O}_{\ell}$ which performs the unitary $\mathcal{O}_{\ell}|x\rangle=(-1)^{\ell(x)}|x\rangle$. In contrast, any classical algorithm with access to a classical oracle computing $\ell$ requires $n$ queries to obtain $z$. The optimal classical strategy is to evaluate $\ell$ for each input $x$ where one of the bits is set to $1$, and all the other $N-1$ bits are set to $0$. We recall the detailed steps of the BV algorithm summarized below.

\begin{itemize}
\item [1.] Apply Hadamard gates $H$ to all qubits (the $n$-qubit input state $|0^n\rangle$ in the first register and the ancilla qubit $|1\rangle_{q}$ in the second register). From the first register we obtain the equal superposition state
\begin{equation}
   |\eta\rangle:=\frac{1}{\sqrt{N}} \sum_{x=0}^{N-1}|x\rangle.
\end{equation}
The system state is $|\eta\rangle|-\rangle_{q}$, where $|-\rangle_{q}=\frac{1}{\sqrt{2}}(|0\rangle-|1\rangle)$.

\item [2.] Perform the oracle denoted as the unitary $\mathcal{O}_{\ell}$: $\mathcal{O}_{\ell}|x\rangle|-\rangle_{q}=(-1)^{\ell(x)}|x\rangle|-\rangle_{q}$. 
Since the state of the ancilla remains unchanged, it is conventional to omit it and represent the action of the oracle as
\begin{equation}\label{O_l}
    \mathcal{O}_{\ell}|x\rangle=(-1)^{\ell(x)}|x\rangle.
\end{equation}

\item [3.] Apply Hadamard gates $H$ to all qubits again.

\item [4.] Measure the first register. The outcome of the measurement will be the secret bit string $z$, and the algorithm terminates.
\end{itemize}

In the third step of the algorithm, applying Hadamard gates to each qubit obtains a classical state $|z\rangle=H^{\otimes n} \mathcal{O}_{\ell} H^{\otimes n}\left|0^n\right\rangle$. As a result, the BV problem only performs a single query to the oracle. The success probability of the BV algorithm is
\begin{equation}\label{p-bv}
   P_{\text{succ}}^{BV}=|\langle z|H^{\otimes n}\mathcal{O}_{\ell} H^{\otimes n}\left|0^n\right\rangle|^2=1. 
\end{equation}

We next analyze the performance of the BV algorithm for general initial states. In general, the procedure may not perform optimally unless the initial state is $|\eta\rangle$. To capture the performance in the general setup, we assume that the initial state is an arbitrary state generated by applying an arbitrary unitary gate.
\subsection{Generalized Bernstein-Vazirani algorithm}
\begin{figure}[H]
\centering
\includegraphics[width=3in]{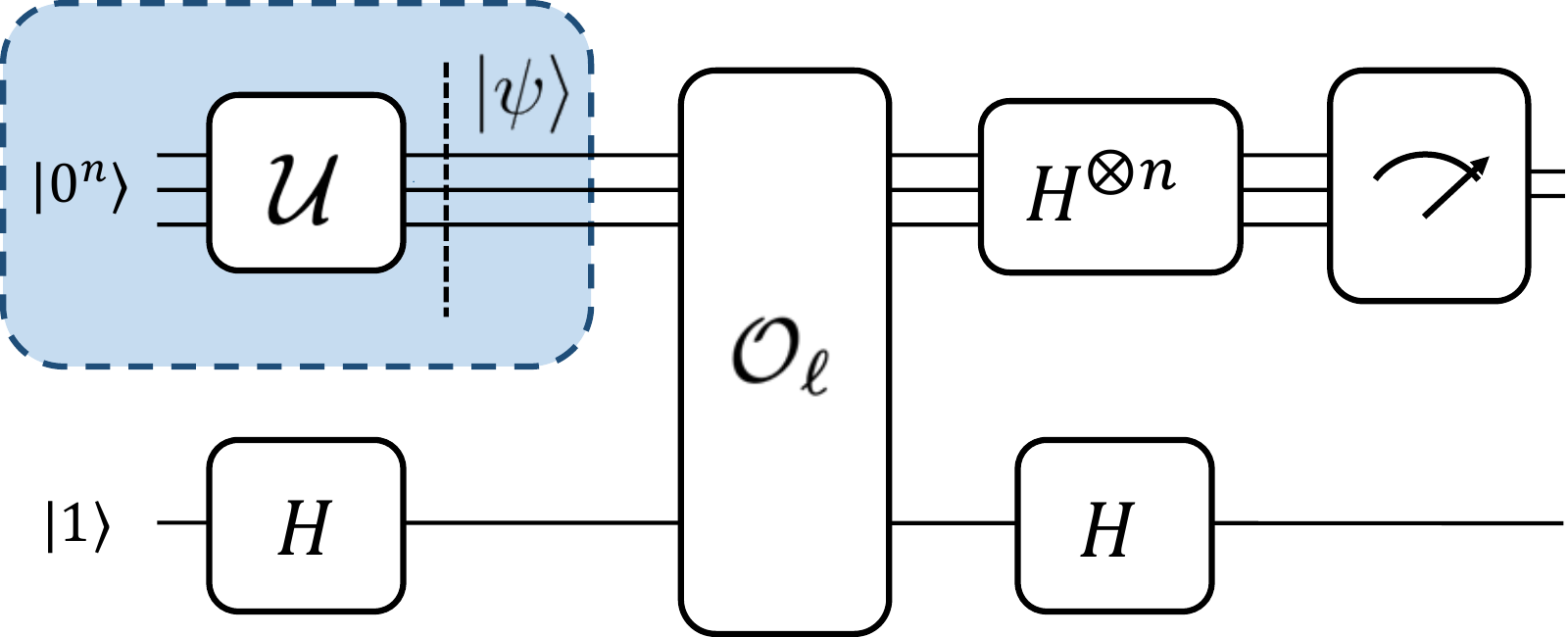}
\caption{Circuit diagram for GBV algorithm. Here two registers with $n+1$ qubits are initialized in the state $|0^n\rangle|1\rangle_{q}$. At first, apply an arbitrary unitary quantum gate $\mathcal{U}$ on the input state $|0^{n}\rangle$ of the first register to obtain an arbitrary initial state $|\psi\rangle$. Apply a Hadamard gate $H$ on the ancilla qubit $|1\rangle_{q}$ in the second register at the same time. Next, perform the oracle $\mathcal{O}_{\ell}$ on the system state and then apply Hadamard gates $H$ to all qubits. Finally, Measure the first register.}
\label{fig:GBV}
\end{figure}

In this section, we present a generalized version of the original BV (GBV) algorithm to study the role of coherence fraction in this algorithm. At first, we consider the case that the initial state of the GBV algorithm is an arbitrary state obtained by using unitary gate $\mathcal{U}$ before the oracle. Based on this idea and the original BV algorithm, we present a GBV algorithm as follows, with its corresponding circuit depicted in FIG.~\ref{fig:GBV}.

\textbf{Algorithm.} \textbf{GBV algorithm.} 

\textbf{Inputs.} (i). Here there are two registers holding $n+1$ qubits in the state $|0^n\rangle|1\rangle_{q}$; (ii). A black box oracle $\mathcal{O}_{\ell}$, whose action is defined by Eq.(\ref{O_l}).

\textbf{Outputs.} The secret bit string $z$.

\textbf{Procedure.}

\textbf{Step 1.} Apply an arbitrary unitary quantum gate $\mathcal{U}$ on the input state $|0^n\rangle$ of the $n$-qubit first register and to obtain an arbitrary state 
\begin{equation}\label{psi}
   |\psi\rangle:=\mathcal{U}|0^n\rangle=\sum_{x=0}^{N-1} a_x|x\rangle, 
\end{equation}
where $a_x$ is the amplitude of the basis state $|x\rangle$. Apply a Hadamard gate $H$ on the ancilla qubit $|1\rangle_{q}$ in the second register at the same time. 

\textbf{Step 2.} Perform the oracle $\mathcal{O}_{\ell}$ defined by Eq.(\ref{O_l}) on the system state.

\textbf{Step 3.} Apply Hadamard gate $H^{\otimes n}$ on the first $n$-qubit register and a Hadamard gate $H$ on the ancilla qubit.

\textbf{Step 4.} Measure the first register. The outcome of the measurement will be the secret bit string $z$, and the GBV algorithm terminates.

The connection is established by inquiring about the probability of success, denoted as $P_{\text{succ}}^{G}$, of uncovering a secret string $z$. This success probability is calculated across all possible arbitrary unitary operations performed during the initialization phase. We show that $P_{\text{succ}}^{G}$ is related to the coherence fraction of the presented initial register state $|\psi\rangle$. The following asserts the success probability of the circuit from FIG.~\ref{fig:GBV} of the GBV algorithm. At first, the success probability of obtaining the output $z\in \{0,1\}^n$ of the algorithm is therefore drawn from 
\begin{equation}\label{succ-0}
P_{\text{succ}}^{G}=|\langle z|H^{\otimes n}\mathcal{O}_l\mathcal{U}|0^n\rangle|^2.   
\end{equation}
Then we investigate the relationship between the success probability of the algorithm and the coherence fraction of the initial state. We finally introduce the coherence fraction dynamics of the states after the operators applied in the GBV algorithm.

\section{Quantum coherence fraction in the GBV algorithm}\label{III}
In this section, we investigate the connections between the success probability of the GBV algorithm and the QCF, covering both state and operator coherence fractions.

\subsection{State coherence fraction of the GBV algorithm}
Following the idea of the coherence fraction of a state~\cite{PhysRevA.100.032324,bai2015maximallycoherentstates,PhysRevA.93.032326,karmakar2019coherence} in the theory of quantum coherence, we present the definition of the state coherence fraction that quantifies the overlap of a state with the equal superposition state (i.e., a maximally coherent state).

\begin{df}[\textbf{State coherence fraction}]
 The coherence fraction of a state $\rho$ is defined as the Uhlmann's fidelity between $\rho$ and $|\eta\rangle $, denoted as
\begin{equation}\label{scf}
      C_{\mathcal{F}}(\rho):= F(|\eta\rangle, \rho)=\langle \eta | \rho |\eta\rangle,  
\end{equation}
where $ |\eta\rangle $ is the maximally coherent state (i.e., the equal superposition state).   
\end{df}
The Uhlmann fidelity of any states $\rho$ and $\sigma$ is generally defined as~\cite{Uhlmann1976The,Jozsa1994Fidelity,Liang2019Quantum}, $F(\rho, \sigma) \equiv [\operatorname{Tr}(\sqrt{\rho} \sigma \sqrt{\rho})^{1 / 2}]^2$. In the special case where one state is pure and the other is mixed, the fidelity simplifies to $F(\sigma, |a\rangle)=\langle a |\sigma| a\rangle$. The coherence fraction quantifies how close a state is to a maximally coherent state. In quantum teleportation, there have been similar results regarding the FEF which measures the optimal fidelity of quantum teleportation. Note that the coherence fraction is neither an entanglement measure nor a coherence measure, but turns out to be a coherence measure under specific conditions. For states $\rho$ such that $\langle i|\rho|j\rangle \geqslant 0$ for all $i$ and $j$, $C(\rho):=C_{\mathcal{F}}(\rho)-1/N$ is just the $l_{1}$-norm of coherence~\cite{PhysRevLett.113.140401}. Therefore, in this case, $C_{\mathcal{F}}(\rho)$ quantifies the coherence, up to a constant factor $1/N$.

Now, we present Theorem \ref{thm1} which connects the coherence fraction of the initial state with the success probability of the GBV algorithm.

\begin{theorem}\label{thm1}
For an initial state $\delta=|\psi\rangle \langle\psi|$, the success probability of the algorithm is given by the following formula,
\begin{equation}\label{succ-1}
P_{\text{succ}}^{G}(\delta)=C_{\mathcal{F}}(\delta).
\end{equation}
where $C_{\mathcal{F}}(\delta)$ represents the coherence fraction of the initial state $\delta$. This coherence fraction is defined by the fidelity $F(|\eta\rangle, \delta)$ between $\delta$ and the maximally coherent state $|\eta\rangle$, where $|\eta\rangle=\sum_{x=0}^{N-1}|x\rangle/\sqrt{N}$.
\end{theorem}

\begin{pf}
According to the success probability of the original BV algorithm in Eq.(\ref{p-bv}), we have the equation $H^{\otimes n}\mathcal{O}_lH^{\otimes n}|0^n\rangle=|z\rangle$. Recall $H^{\otimes n}|0^n\rangle=|\eta\rangle$, then we rewrite the equation as $H^{\otimes n}\mathcal{O}_l|\eta\rangle=|z\rangle$. Thus, we have
\begin{align}
    P_{\text{succ}}^{G}(|\psi\rangle)&=|\langle z|H^{\otimes n}\mathcal{O}_l\mathcal{U}|0^n\rangle|^2\nonumber\\
    &=|\langle \eta|\mathcal{O}_l^{\dag}H^{\dag\otimes n}H^{\otimes n}\mathcal{O}_l|\psi\rangle|^2\nonumber\\
    &=|\langle \eta |\psi\rangle|^2=F(|\eta\rangle, |\psi\rangle).
\end{align}
Obviously, $F(|\eta\rangle, |\psi\rangle)$ corresponds to the coherence fraction of the pure state $|\psi\rangle$ based on the definition in Eq.(\ref{scf}). Extending this result to the initial state $\delta=|\psi\rangle \langle\psi|$, we obtain the relation $P_{\text{succ}}^{G}(\delta)=C_{\mathcal{F}}(\delta)$. Note that a similar analysis can extend to an arbitrary $n$-qubit mixed initial state, $\rho=\sum_{\mu}p_{\mu}|\psi_{\mu}\rangle\langle\psi_{\mu}|$, where $\sum_{\mu}p_{\mu}=1$ and pure state $|\psi_{\mu}\rangle=\sum_{i=0}^{N-1}a_{\mu i}|i\rangle$ with $\sum_{i=0}^{N-1}|a_{\mu i}|^2=1$.
\qed
\end{pf}

According to Eq.(\ref{succ-1}), the success probability of the GBV algorithm is fully determined by the coherence fraction of the initial state.

\subsection{Operator coherence fraction of the GBV algorithm}
After introducing the definition of the state coherence fraction, we are in a position to introduce another information theoretic quantifier. Concerning the established definition of operator coherence~\cite{PhysRevA.100.012349}, we present a corresponding definition for the operator coherence fraction to clarify the characteristics of the coherence fraction of the state after each operator is applied in the algorithm.

\begin{df}[\textbf{Operator coherence fraction}]
    Let $U$ be a unitary operator. Operator coherence fraction of $U$ with respect to a state $|\phi \rangle$ is the coherence fraction of the state after $U$ operating on $|\phi\rangle$ that 
    \begin{equation}\label{ocf}
     C_{\mathcal{F}}(\rho^{U})=C_{\mathcal{F}}(U \rho U^{\dagger}),
    \end{equation}
    where $\rho=|\phi\rangle \langle \phi|$. 
\end{df}

The operator coherence fraction quantifies the coherence of an operator with respect to a specific reference state. Different reference states may lead to different coherence fractions, revealing how the operator behaves in various quantum environments. Since the operator coherence fraction depends on the chosen reference state, it can exhibit extreme behaviors under certain conditions. For example, selecting a specific reference state might make the operator’s coherence fraction approach zero or one, or in some cases, the operator coherence fraction can theoretically take any value. This phenomenon reflects the flexibility of the coherence measure in the context of different quantum states and operators.

Studying the operator coherence fraction helps us understand how operators affect the coherence properties of quantum states, which is crucial in quantum information processing and algorithms. Specifically, it can reveal how operators preserve or disrupt coherence, which is important for understanding quantum speedup and other key areas in quantum computing. In this regard, the operator coherence fraction provides key insights into the efficiency and stability of quantum algorithms. Although this work focuses on unitary operators due to their importance in quantum computing, the concept of the operator coherence fraction is not limited to unitary operations. It can be extended to more general quantum channels, including open quantum systems and noisy channels~\cite{karmakar2019coherence}, thereby broadening its applicability to a wider range of quantum scenarios.

Now, we investigate the coherence fraction dynamics of the three operators $-$ an arbitrary unitary quantum operator $\mathcal{U}$, the oracle $\mathcal{O}_{\ell}$, and Hadamard gate $H^{\otimes n}$ $-$ applied in the GBV algorithm. We establish a direct link between the coherence fraction of each operator and the success probability of the algorithm at the same time.

\begin{theorem}\label{thm2}
The coherence fraction of an arbitrary unitary quantum operator $\mathcal{U}$ with respect to the input state $\rho^0$ is given by
\begin{equation}\label{CFU}
     C_{\mathcal{F}}(\rho^{\mathcal{U}})=P_{\text{succ}}^{G}(\delta),
\end{equation}
where $\rho^{\mathcal{U}}=\mathcal{U}\rho^0 \mathcal{U}^{\dagger}$, $\rho^0=|0^n\rangle \langle 0^n|$ is the input state and $\delta=|\psi\rangle \langle\psi|$ is the initial state of the GBV algorithm. 
\end{theorem}

\begin{pf}
From the first step of the GBV algorithm, an arbitrary unitary quantum gate $\mathcal{U}$ transforms the input state $|0^n\rangle$ into an arbitrary initial state $|\psi\rangle$. Thus the state after the operator $\mathcal{U}$ is 
\begin{equation}\label{rhou}
\begin{aligned}
    \rho^{\mathcal{U}}&=\mathcal{U}\rho^0 \mathcal{U}^{\dagger}=\mathcal{U}|0^n\rangle \langle 0^n|\mathcal{U}^{\dagger}=|\psi\rangle \langle \psi|\\
    &=\sum_{x,y=0}^{N-1}a_xa_y^*|x\rangle \langle y|,
\end{aligned}
\end{equation}
where $|\psi\rangle=\mathcal{U}|0^n\rangle=\sum_{x=0}^{N-1} a_x|x\rangle$ is as shown in Eq.(\ref{psi}), $a_x$ $(a_y)$ is the amplitude of the basis state $|x\rangle$ $(|y\rangle)$ and $a^*$ is the conjugation of the complex number $a$. Based on the state coherence fraction defined in Eq.(\ref{scf}) and the operator coherence fraction in Eq.(\ref{ocf}), we have 
\begin{equation}
     C_{\mathcal{F}}(\rho^{\mathcal{U}})=\langle \eta | \rho^{\mathcal{U}} |\eta\rangle=\frac{1}{N}|\sum_{x=0}^{N-1}a_x|^2=P_{\text{succ}}^{G}(\delta).
\end{equation}
It is easy to see that the state after applying an arbitrary unitary quantum gate $\mathcal{U}$ is the initial state $|\psi\rangle$ of the GBV algorithm. This indicates that $\rho^{\mathcal{U}}=\delta=|\psi\rangle \langle\psi|$. Consequently, the coherence fraction $C_{\mathcal{F}}(\rho^{\mathcal{U}})$ can be directly calculated as $P_{\text{succ}}^{G}(\delta)$, according to Theorem \ref{thm1}.
\qed
\end{pf}

From Eq.(\ref{CFU}), it is evident that the coherence fraction of an arbitrary unitary quantum operator $\mathcal{U}$ equals the success probability of the GBV algorithm. Moreover, we demonstrate that the Hadamard operator $H$ achieves the maximum coherence fraction, which is equal to $1$. Our findings elucidate why the Hadamard gate is frequently used in the BV algorithm, as opposed to an arbitrary quantum unitary gate.

\begin{theorem}\label{thm3}
    The coherence fraction of the oracle $\mathcal{O}_{\ell}$ with respect to the state $\rho^{\mathcal{U}}$ can be expressed as
    \begin{equation}
     C_{\mathcal{F}}(\rho^{\mathcal{O}_{\ell}})=P_{\text{succ}}^{G}(\delta)-\frac{4}{N}\Re(\sum_{s\in S, t\in T}a_sa_t^*),
    \end{equation}
where $\rho^{\mathcal{O}_{\ell}}=\mathcal{O}_{\ell} \rho^{\mathcal{U}} \mathcal{O}_{\ell}^{\dagger}$ and $\rho^{\mathcal{U}}=|\psi\rangle \langle \psi|$ as shown in Eq.(\ref{rhou}) of the Theorem \ref{thm2}. Here, $a_s$ and $a_t$ are the amplitudes of the basis states such that $\ell(s)=0$ and $\ell(t)=1$, respectively. And $\Re(\sum_{s,t}a_sa_t^*)$ is the real part of $\sum_{s,t}a_sa_t^*$.
\end{theorem}

\begin{pf}
   From the step 2 of the GBV algorithm, we perform the oracle $\mathcal{O}_{\ell}$ on the initial state $|\psi\rangle$ to get 
\begin{equation}
    |\psi^{\mathcal{O}_{\ell}}\rangle=\mathcal{O}_{\ell}|\psi\rangle=\sum_{x=0}^{N-1}(-1)^{\ell(x)} a_x|x\rangle.
\end{equation} 
Its density matrix is 
\begin{equation}\label{rhoo}
    \rho^{\mathcal{O}_{\ell}}=|\psi^{\mathcal{O}_{\ell}}\rangle \langle\psi^{\mathcal{O}_{\ell}}|=\sum_{x,y=0}^{N-1}(-1)^{\ell(x)+\ell(y)} a_xa_y^*|x\rangle \langle y|.
\end{equation}
According to the definition of the state coherence fraction, we have
\begin{equation}
     C_{\mathcal{F}}(\rho^{\mathcal{O}_{\ell}})=\langle \eta | \rho^{\mathcal{O}_{\ell}} |\eta\rangle=\frac{1}{N}|\sum_{x=0}^{N-1}(-1)^{\ell(x)} a_x|^2.
\end{equation}
We know that $\ell: \mathbb{F}_{2}^{n} \rightarrow \mathbb{F}_{2}$ is a linear boolean function, which holds $\ell(x)\in\{0, 1\}$. Denote the set $S$ as the collection of indices associated with the basis states where $\ell(x)=0$, and let the corresponding amplitudes be $a_s$, for $s\in S$. The complementary set $T$ consists of the indices for the basis states where $\ell(x)=1$, with corresponding amplitudes $a_t$, for $t\in T$. Thus, the operator coherence fraction of the $\mathcal{O}_{\ell}$ can be re-expressed by
\begin{equation}\label{st}
     C_{\mathcal{F}}(\rho^{\mathcal{O}_{\ell}})=\frac{1}{N}|\sum_{s\in S}a_s-\sum_{t\in T}a_t|^2.
\end{equation}
Based on the results in Theorem \ref{thm1} and Theorem \ref{thm2}, we have 
\begin{equation}
     P_{\text{succ}}^{G}(\delta)=\frac{1}{N}|\sum_{x=0}^{N-1}a_x|^2=\frac{1}{N}|\sum_{s\in S}a_s+\sum_{t\in T}a_t|^2.
\end{equation}
Thus we can rewrite the Eq.(\ref{st}) as
\begin{equation}
     C_{\mathcal{F}}(\rho^{\mathcal{O}_{\ell}})=P_{\text{succ}}^{G}(\delta)-\frac{4}{N}\Re(\sum_{s\in S, t\in T}a_sa_t^*),
\end{equation}
where $\Re(\sum_{s,t}a_sa_t^*)$ is the real part of $\sum_{s,t}a_sa_t^*$. 
\qed
\end{pf}

From Theorem~\ref{thm3}, we find that the operator coherence fraction of the $\mathcal{O}_{\ell}$ depends not only on the oracle queries but also on the initial state. In particular, $C_{\mathcal{F}}(\rho^{\mathcal{O}_{\ell}})=C_{\mathcal{F}}(\rho^{\mathcal{U}})-\frac{4}{N}\Re(\sum_{s\in S, t\in T}a_sa_t^*)$,
which implies that the oracle would change or not change the coherence fraction of the state $\rho^{\mathcal{U}}$.

Now, we show the coherence fraction after performing the operator $H^{\otimes n}$.

\begin{theorem}\label{thm4}
    The coherence fraction of the Hadamard operator $H^{\otimes n}$ with respect to the state $\rho^{\mathcal{O}_{\ell}}$ is given by
    \begin{equation}\label{rhoh}
     C_{\mathcal{F}}(\rho^{H^{\otimes n}})=\frac{1}{N}P_{\text{succ}}^{G}(\delta),
    \end{equation}
where $\rho^{H^{\otimes n}}=H^{\otimes n} \rho^{\mathcal{O}_{\ell}} H^{\otimes n}$ and $\rho^{\mathcal{O}_{\ell}}$ is represented in Eq.(\ref{rhoo}) in the Theorem \ref{thm3}. 
\end{theorem}

\begin{pf}
Remember that the Hadamard transform may be defined using the bit-wise dot product $x\cdot y$ as: $H^{\otimes n}|x\rangle=\frac{1}{\sqrt{N}}\sum_{y=0}^{N-1}(-1)^{y\cdot x}|y\rangle$. Using this notation, the result of applying Hadamard operations is
\begin{equation}
\begin{aligned}
|\psi^{H^{\otimes n}}\rangle&=H^{\otimes n}|\psi^{\mathcal{O}_{\ell}}\rangle=H^{\otimes n}\sum_{x=0}^{N-1}(-1)^{z\cdot x} a_x|x\rangle\\
&=\frac{1}{\sqrt{N}}\sum_{x,y=0}^{N-1}(-1)^{(z\oplus y)\cdot x}a_{x}|y\rangle,
\end{aligned}
\end{equation}
where $|\psi^{\mathcal{O}_{\ell}}\rangle=\sum_{x=0}^{N-1}(-1)^{\ell(x)} a_x|x\rangle$, and the linear boolean function $\ell$ parameterized by a secret bit string $z\in\{0,1\}^{n}$, such that $\ell(x)=z\cdot x\bmod2$, $x\in\{0,1\}^{n}$. 
It is important to note that when $y \neq z$, the success probability of the GBV algorithm $|\langle z|\psi^{H^{\otimes n}}\rangle|^2=\frac{1}{N}|\sum_{x,y=0}^{N-1}(-1)^{(z\oplus y)\cdot x}a_{x}\langle z|y\rangle|^2=0$. Therefore, we only need to consider the case $y=z$. Then the state is
\begin{equation}
|\psi^{H^{\otimes n}}\rangle=\frac{1}{\sqrt{N}}\sum_{x=0}^{N-1}a_{x}|z\rangle,
\end{equation}
Thus its density matrix is
\begin{equation}
\rho^{H^{\otimes n}}=|\psi^{H^{\otimes n}}\rangle \langle\psi^{H^{\otimes n}}|=\frac{1}{N}|\sum_{x=0}^{N-1}a_{x}|^2|z\rangle\langle z|.
\end{equation}
Then the coherence fraction of the Hadamard operator $H^{\otimes n}$ is given by
\begin{equation}
     C_{\mathcal{F}}(\rho^{H^{\otimes n}})=\langle\eta|\rho^{H^{\otimes n}}|\eta\rangle=\frac{1}{N^2}|\sum_{x}^{N-1}a_{x}|^2=\frac{1}{N}P_{\text{succ}}^{G}(\delta).
\end{equation}
This completes the proof of Theorem \ref{thm4}. 
\qed
\end{pf}
Combined with Eq.(\ref{rhoh}) and Theorem~\ref{thm2}, we find that $C_{\mathcal{F}}(\rho^{H^{\otimes n}}) \le C_{\mathcal{F}}(\rho^{0})=1/N$, and $C_{\mathcal{F}}(\rho^{H^{\otimes n}}) \le C_{\mathcal{F}}(\rho^{\mathcal{U}})$.

\section{Example}\label{IV}
Let us consider an example to illustrate the coherence fraction in the GBV algorithm. We consider a special case in which the unitary quantum gate we applied in the GBV algorithm is a product of arbitrary local operations. The unitary gate applied to each qubit in the register is
\begin{equation}
   \mathcal{U}=\Big(\mathcal{U}(\alpha, \beta, \theta)\Big)^{\otimes{n}},
\end{equation}
where the unitary gate $\mathcal{U}(\alpha, \beta, \theta)$ applied to each qubit in the first register is given by
\begin{equation}
\mathcal{U}(\alpha, \beta, \theta)
=\begin{bmatrix}
    e^{i\alpha}\cos\theta & e^{-i\beta}\sin\theta \\
    e^{i\beta}\sin\theta & -e^{-i\alpha}\cos\theta
\end{bmatrix}
\end{equation}
with parameters $\alpha$, $\beta$ and $\theta \in [0, \pi/2]$. 

Applying the unitary gate to the single qubit state $|0\rangle$, we obtain
\begin{equation}
\begin{aligned}
|\phi(\alpha, \beta, \theta)\rangle&=\mathcal{U}(\alpha, \beta, \theta)|0\rangle \\
&=e^{i\alpha}\cos\theta|0\rangle+e^{i\beta}\sin\theta|1\rangle.
\end{aligned}
\end{equation}
The initial state is then given by
\begin{equation}\label{psialpha}
\begin{aligned}
    |\psi(\alpha, \beta, \theta)\rangle&=|\phi(\alpha, \beta, \theta)\rangle^{\otimes n}\\
    &=\sum_{x=0}^{N-1}(e^{i\alpha}\cos\theta)^{n-H(x)}(e^{i\beta}\sin\theta)^{H(x)}|x\rangle,
\end{aligned}
\end{equation}
where $N=2^n$ and $H(x)$ is the Hamming weight (number of $1$'s) of the binary representation of $x=x_{1}x_{2}\cdots x_{n}$.

\begin{figure*}
    \centering
\includegraphics[width=6in]{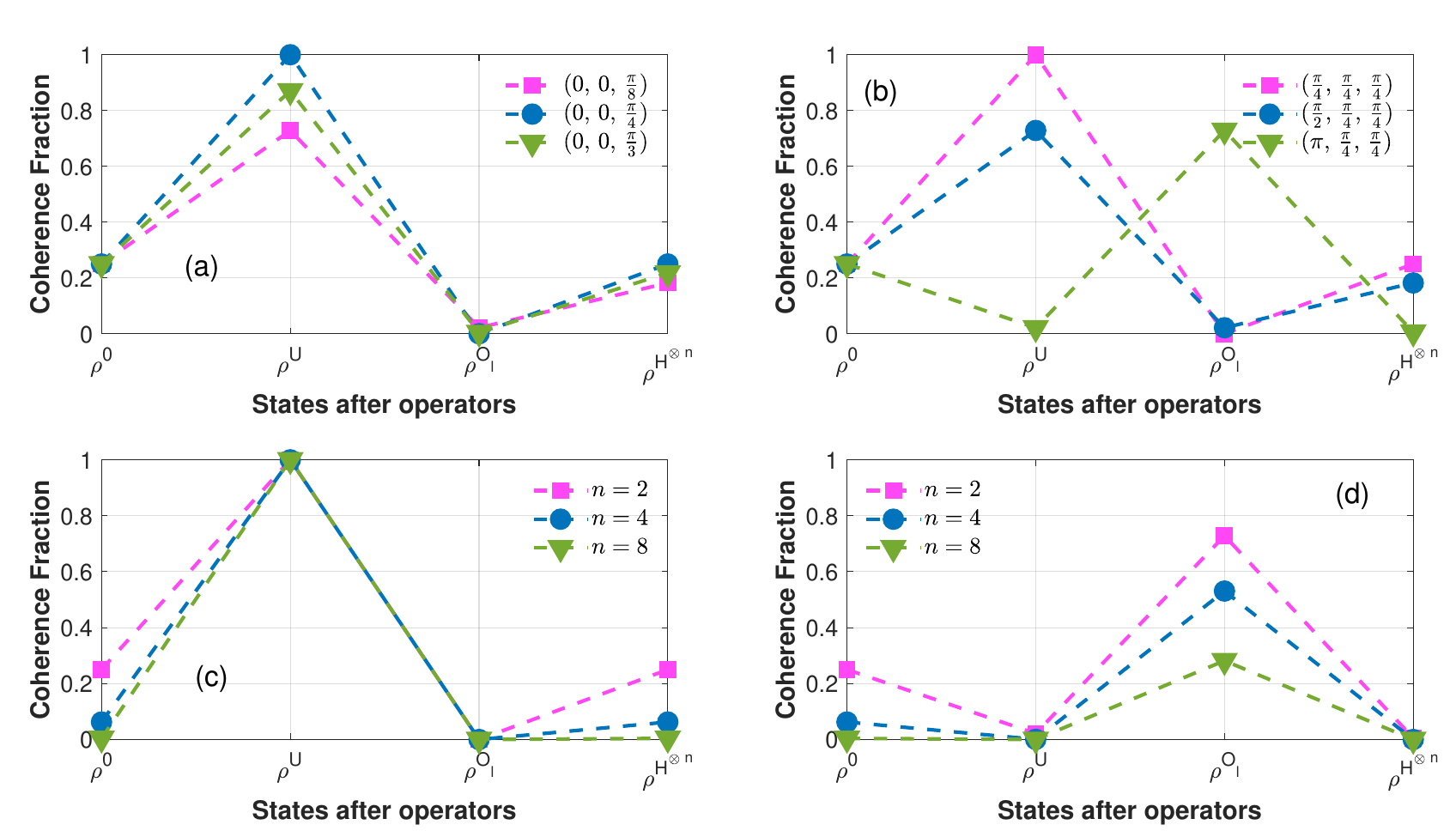}
\caption{\textbf{The coherence fraction dynamics of the GBV algorithm.} The link between $n$, parameters $(\alpha, \beta, \theta)$ and the coherence fraction of the states after three operators applied in the algorithm. \textbf{(a).} For a fixed $n=2$ and parameters $\alpha=\beta=0$, the three lines in the sub-figure correspond to $\theta=\pi/8$, $\pi/4$, and $\pi/3$, respectively. \textbf{(b).} For $n=2$, with parameters $\beta=\theta=\pi/4$, the sub-figure shows three lines for $\alpha=\pi/4$, $\pi/2$, and $\pi$. \textbf{(c).} With parameters $\alpha=\beta=\theta=\pi/4$, the three lines in the sub-figure demonstrate $n=2, 4$, and $8$, respectively. \textbf{(d).} With $\alpha=\pi$ and $\beta=\theta=\pi/4$, the sub-figure shows three lines corresponding to $n=2, 4$, and $8$.}
\label{fig:Fig}
\end{figure*}

The coherence fraction of the initial state $\delta(\alpha, \beta, \theta)=|\psi(\alpha, \beta, \theta)\rangle \langle\psi(\alpha, \beta, \theta)|$ is
\begin{equation}
C_{\mathcal{F}}\Big(\delta(\alpha, \beta, \theta)\Big)=\frac{1}{2^{n}}|(e^{i\alpha}\cos\theta+e^{i\beta}\sin\theta)^{n}|^{2}.
\end{equation}
The success probability of the GBV algorithm is given by
\begin{equation}
\begin{aligned}
P_{\text{succ}}^{G}\Big(\delta(\alpha, \beta, \theta)\Big)
&=C_{\mathcal{F}}\Big(\delta(\alpha, \beta, \theta)\Big)\\
&=\frac{1}{2^{n}}|(e^{i\alpha}\cos\theta+e^{i\beta}\sin\theta)^{n}|^{2}.
\end{aligned}
\end{equation}
Then, we investigate the coherence fraction dynamics of the states after the three operators: (i) an arbitrary unitary quantum operator $\mathcal{U}$, (ii) the oracle $\mathcal{O}_{\ell}$, (iii) and Hadamard gate $H^{\otimes n}$. 

\textbf{(i).} The coherence fraction of the arbitrary unitary quantum operator $\mathcal{U}(\alpha, \beta, \theta)$ is given by
\begin{equation}
     C_{\mathcal{F}}(\rho^{\mathcal{U}})=\frac{1}{2^{n}}|(e^{i\alpha}\cos\theta+e^{i\beta}\sin\theta)^{n}|^{2}.
\end{equation}

\textbf{(ii).} From the Theorem \ref{thm3}, the coherence fraction of the oracle $\mathcal{O}_{\ell}$ depends on the oracle and the initial state. Thus, based on the initial state in Eq.(\ref{psialpha}) and the oracle $\mathcal{O}_{\ell}|x\rangle=(-1)^{\ell(x)}|x\rangle$, we have 
    \begin{equation}\label{rhooraex}
    \begin{aligned}
       C_{\mathcal{F}}(\rho^{\mathcal{O}_{\ell}})=&\frac{1}{2^n}|\sum_{s\in S}(e^{i\alpha}\cos\theta)^{n-H(s)}(e^{i\beta}\sin\theta)^{H(s)}\\
       &-\sum_{t\in T}(e^{i\alpha}\cos\theta)^{n-H(t)}(e^{i\beta}\sin\theta)^{H(t)}|^2, 
    \end{aligned}
    \end{equation}
where $s\in S$ such that $\ell(s)=0$ and $t\in T$ such that $\ell(t)=1$. Here, we consider a special case in which the relation for bit strings and the linear boolean function are specified by $\ell(s)=0$ if $H(s)$ is even, and $\ell(t)=1$ if $H(t)$ is odd. Thus, the Eq.(\ref{rhooraex}) will change to 
\begin{equation}
     C_{\mathcal{F}}(\rho^{\mathcal{O}_{\ell}})=\frac{1}{2^n}|(e^{i\alpha}\cos\theta-e^{i\beta}\sin\theta)^{n}|^{2}.
\end{equation}

\textbf{(iii).} The coherence fraction of the Hadamard operator $H^{\otimes n}$ is given by
\begin{equation}
\begin{aligned}
C_{\mathcal{F}}(\rho^{H^{\otimes n}})&=\frac{1}{N}P_{\text{succ}}^{G}\Big(\delta(\alpha, \beta, \theta)\Big) \\
&=\frac{1}{4^{n}}|(e^{i\alpha}\cos\theta+e^{i\beta}\sin\theta)^{n}|^{2}.
\end{aligned}
\end{equation}

The coherence fraction dynamics of the states in this example are shown in FIG. \ref{fig:Fig}. We explore the link between the parameters $(\alpha, \beta, \theta)$ of the initial state and the coherence fraction of the states after three operators are applied in the GBV algorithm. We also present the dynamics of the coherence fraction as it varies with $n$. From FIG. \ref{fig:Fig}, we observe that $C_{\mathcal{F}}(\rho^{0})$ is independent of the parameters (from \textbf{(a)} and \textbf{(b)}) and is solely determined by $n$, decreasing as $n$ increases (from \textbf{(c)} and \textbf{(d)}). According to the two sub-figures \textbf{(a)} and \textbf{(b)}, we also observe that the maximum value of $C_{\mathcal{F}}(\rho^{\mathcal{U}})=1$ is attained when the parameters satisfy $\alpha=\beta+2k\pi$ (where $k=0, \pm 1, \cdots$) and $\theta=\pi/4$, which corresponds to the case where the operator $\mathcal{U}$ is the Hadamard operator. Sub-figure \textbf{(c)} shows that when $\alpha=\beta=\theta=\pi/4$, $C_{\mathcal{F}}(\rho^{\mathcal{U}})$ and $C_{\mathcal{F}}(\rho^{\mathcal{O}_{\ell}})$ are independent of $n$. Combined with the four sub-figures, we find that $C_{\mathcal{F}}(\rho^{\mathcal{O}_{\ell}})$ can be greater than $C_{\mathcal{F}}(\rho^{\mathcal{U}})$, as well as smaller than it, depending on the choice of parameters.

\section{Discussions}\label{V}
In this work, we have introduced the concept of the state and operator coherence fraction, which serves as an analog to the fully entangled fraction in entanglement theory, to quantify how close a quantum state is to a maximally coherent state. By applying this quantity to the Bernstein-Vazirani algorithm, we explored the connections between the success probability of a generalized Bernstein-Vazirani algorithm and the coherence fraction of both state and operator. Our results emphasize that quantum coherence fraction plays a pivotal role in optimizing the performance of quantum algorithms, extending the utility of coherence beyond traditional entanglement-based measures. This study paves the way for further investigations into the interplay between entanglement and coherence in quantum information processing, with potential implications for a wide range of quantum algorithms and protocols. 

In literature~\cite{duan2001long}, examples show that communication efficiency scales polynomially with channel length, a phenomenon closely tied to the entanglement fraction. It is expected that, with proper implementation, the coherence fraction will play a similarly pivotal role in determining the effectiveness of various experimental or communication schemes. For instance, in entanglement swapping, the efficient conversion of atomic excitations into photons is crucial, and this efficiency can be directly linked to coherence fraction. This suggests that the coherence fraction, much like the entanglement fraction, serves as a key metric for optimizing and evaluating the performance of quantum communication and experimental schemes.

Shor's algorithm~\cite{365700}, addresses the integer factorization and discrete logarithm problems, which are intractable for classical computers. Although Shor's algorithm and the Bernstein-Vazirani algorithm tackle distinct problems, they are both rooted in fundamental quantum computing principles. Specifically, they leverage quantum superposition, interference, and the quantum Fourier transform to achieve exponentially improvements in computational efficiency and reductions in time complexity. Given this shared foundation, further exploration of the role of the coherence fraction in Shor’s algorithm could provide valuable insights into optimizing its performance and enhancing our understanding of its underlying mechanisms. Such investigations may also shed light on the broader relationship between coherence and quantum computational advantages.

\section*{Acknowledgements}
This work was supported by Fundamental Research Funds for the Central Universities, the National Natural Science Foundation of China (12371132, 12075159, 12071179, 12171044, 12405006, 11071178), the Postdoctoral Fellowship Program of CPSF (No. GZC20230103), the China Postdoctoral Science Foundation (No. 2023M740118), and the specific research fund of the Innovation Platform for Academicians of Hainan Province.
\bibliography{bv}

@article{PhysRevLett.78.2275,
  title = {Quantifying Entanglement},
  author = {Vedral, V. and Plenio, M. B. and Rippin, M. A. and Knight, P. L.},
  journal = {Phys. Rev. Lett.},
  volume = {78},
  issue = {12},
  pages = {2275--2279},
  numpages = {0},
  year = {1997},
  month = {Mar},
  publisher = {American Physical Society},
  doi = {10.1103/PhysRevLett.78.2275},
  url = {https://link.aps.org/doi/10.1103/PhysRevLett.78.2275}
  }

@article{RevModPhys.81.865,
  title = {Quantum entanglement},
  author = {Horodecki, Ryszard and Horodecki, Pawe\l{} and Horodecki, Micha\l{} and Horodecki, Karol},
  journal = {Rev. Mod. Phys.},
  volume = {81},
  issue = {2},
  pages = {865--942},
  numpages = {0},
  year = {2009},
  month = {Jun},
  publisher = {American Physical Society},
  doi = {10.1103/RevModPhys.81.865},
  url = {https://link.aps.org/doi/10.1103/RevModPhys.81.865}
}

@article{PhysRevLett.69.2881,
  title = {Communication via one- and two-particle operators on Einstein-Podolsky-Rosen states},
  author = {Bennett, Charles H. and Wiesner, Stephen J.},
  journal = {Phys. Rev. Lett.},
  volume = {69},
  issue = {20},
  pages = {2881--2884},
  numpages = {0},
  year = {1992},
  month = {Nov},
  publisher = {American Physical Society},
  doi = {10.1103/PhysRevLett.69.2881},
  url = {https://link.aps.org/doi/10.1103/PhysRevLett.69.2881}
}

@article{PhysRevLett.87.077902,
  title = {Remote State Preparation},
  author = {Bennett, Charles H. and DiVincenzo, David P. and Shor, Peter W. and Smolin, John A. and Terhal, Barbara M. and Wootters, William K.},
  journal = {Phys. Rev. Lett.},
  volume = {87},
  issue = {7},
  pages = {077902},
  numpages = {4},
  year = {2001},
  month = {Jul},
  publisher = {American Physical Society},
  doi = {10.1103/PhysRevLett.87.077902},
  url = {https://link.aps.org/doi/10.1103/PhysRevLett.87.077902}
}

@article{PhysRevLett.87.047901,
  title = {Good Dynamics versus Bad Kinematics: Is Entanglement Needed for Quantum Computation?},
  author = {Linden, Noah and Popescu, Sandu},
  journal = {Phys. Rev. Lett.},
  volume = {87},
  issue = {4},
  pages = {047901},
  numpages = {4},
  year = {2001},
  month = {Jul},
  publisher = {American Physical Society},
  doi = {10.1103/PhysRevLett.87.047901},
  url = {https://link.aps.org/doi/10.1103/PhysRevLett.87.047901}
}

@article{PhysRevLett.110.060504,
  title = {Universal Quantum Computation with Little Entanglement},
  author = {Van den Nest, Maarten},
  journal = {Phys. Rev. Lett.},
  volume = {110},
  issue = {6},
  pages = {060504},
  numpages = {4},
  year = {2013},
  month = {Feb},
  publisher = {American Physical Society},
  doi = {10.1103/PhysRevLett.110.060504},
  url = {https://link.aps.org/doi/10.1103/PhysRevLett.110.060504}
}

@article{PhysRevX.5.041008,
  title = {Resource Theory of Steering},
  author = {Gallego, Rodrigo and Aolita, Leandro},
  journal = {Phys. Rev. X},
  volume = {5},
  issue = {4},
  pages = {041008},
  numpages = {20},
  year = {2015},
  month = {Oct},
  publisher = {American Physical Society},
  doi = {10.1103/PhysRevX.5.041008},
  url = {https://link.aps.org/doi/10.1103/PhysRevX.5.041008}
}

@article{giovannetti2004quantum,
  title={Quantum-enhanced measurements: beating the standard quantum limit},
  author={Giovannetti, Vittorio and Lloyd, Seth and Maccone, Lorenzo},
  journal={Science},
  volume={306},
  number={5700},
  pages={1330--1336},
  year={2004},
  publisher={American Association for the Advancement of Science},
  doi={10.1126/science.1104149}
}

@article{giovannetti2011advances,
  title={Advances in quantum metrology},
  author={Giovannetti, Vittorio and Lloyd, Seth and Maccone, Lorenzo},
  journal={Nat. Photon.},
  volume={5},
  number={4},
  pages={222--229},
  year={2011},
  publisher={Nature Publishing Group UK London},
  doi={https://doi.org/10.1038/nphoton.2011.35}
}

@article{PhysRevX.5.021001,
  title = {Quantum Coherence, Time-Translation Symmetry, and Thermodynamics},
  author = {Lostaglio, Matteo and Korzekwa, Kamil and Jennings, David and Rudolph, Terry},
  journal = {Phys. Rev. X},
  volume = {5},
  issue = {2},
  pages = {021001},
  numpages = {11},
  year = {2015},
  month = {Apr},
  publisher = {American Physical Society},
  doi = {10.1103/PhysRevX.5.021001},
  url = {https://link.aps.org/doi/10.1103/PhysRevX.5.021001}
}

@article{lostaglio2015description,
  title={Description of quantum coherence in thermodynamic processes requires constraints beyond free energy},
  author={Lostaglio, Matteo and Jennings, David and Rudolph, Terry},
  journal={Nat. Commun.},
  volume={6},
  number={1},
  pages={6383},
  year={2015},
  publisher={Nature Publishing Group UK London},
  doi={https://doi.org/10.1038/ncomms7383}
}

@article{PhysRevLett.113.140401,
  title = {Quantifying Coherence},
  author = {Baumgratz, T. and Cramer, M. and Plenio, M. B.},
  journal = {Phys. Rev. Lett.},
  volume = {113},
  issue = {14},
  pages = {140401},
  numpages = {5},
  year = {2014},
  month = {Sep},
  publisher = {American Physical Society},
  doi = {10.1103/PhysRevLett.113.140401},
  url = {https://link.aps.org/doi/10.1103/PhysRevLett.113.140401}
}

@article{PhysRevLett.115.020403,
  title = {Measuring Quantum Coherence with Entanglement},
  author = {Streltsov, Alexander and Singh, Uttam and Dhar, Himadri Shekhar and Bera, Manabendra Nath and Adesso, Gerardo},
  journal = {Phys. Rev. Lett.},
  volume = {115},
  issue = {2},
  pages = {020403},
  numpages = {6},
  year = {2015},
  month = {Jul},
  publisher = {American Physical Society},
  doi = {10.1103/PhysRevLett.115.020403},
  url = {https://link.aps.org/doi/10.1103/PhysRevLett.115.020403}
}

@article{du2015coherencemeasuresoptimalconversion,
author = {Du, Shuanping and Bai, Zhaofang and Qi, Xiaofei},
title = {Coherence measures and optimal conversion for coherent states},
year = {2015},
issue_date = {November 2015},
publisher = {Rinton Press, Incorporated},
address = {Paramus, NJ},
volume = {15},
number = {15–16},
issn = {1533-7146},
journal = {Quantum Info. Comput.},
month = nov,
pages = {1307–1316},
numpages = {10},
doi={
https://doi.org/10.48550/arXiv.1504.02862}
}

@article{PhysRevLett.116.080402,
  title = {Converting Nonclassicality into Entanglement},
  author = {Killoran, N. and Steinhoff, F. E. S. and Plenio, M. B.},
  journal = {Phys. Rev. Lett.},
  volume = {116},
  issue = {8},
  pages = {080402},
  numpages = {6},
  year = {2016},
  month = {Feb},
  publisher = {American Physical Society},
  doi = {10.1103/PhysRevLett.116.080402},
  url = {https://link.aps.org/doi/10.1103/PhysRevLett.116.080402}
}

@article{PhysRevLett.116.120404,
  title = {Operational Resource Theory of Coherence},
  author = {Winter, Andreas and Yang, Dong},
  journal = {Phys. Rev. Lett.},
  volume = {116},
  issue = {12},
  pages = {120404},
  numpages = {6},
  year = {2016},
  month = {Mar},
  publisher = {American Physical Society},
  doi = {10.1103/PhysRevLett.116.120404},
  url = {https://link.aps.org/doi/10.1103/PhysRevLett.116.120404}
}

@article{PhysRevLett.116.160407,
  title = {Converting Coherence to Quantum Correlations},
  author = {Ma, Jiajun and Yadin, Benjamin and Girolami, Davide and Vedral, Vlatko and Gu, Mile},
  journal = {Phys. Rev. Lett.},
  volume = {116},
  issue = {16},
  pages = {160407},
  numpages = {5},
  year = {2016},
  month = {Apr},
  publisher = {American Physical Society},
  doi = {10.1103/PhysRevLett.116.160407},
  url = {https://link.aps.org/doi/10.1103/PhysRevLett.116.160407}
}

@article{PhysRevLett.116.150502,
  title = {Robustness of Coherence: An Operational and Observable Measure of Quantum Coherence},
  author = {Napoli, Carmine and Bromley, Thomas R. and Cianciaruso, Marco and Piani, Marco and Johnston, Nathaniel and Adesso, Gerardo},
  journal = {Phys. Rev. Lett.},
  volume = {116},
  issue = {15},
  pages = {150502},
  numpages = {6},
  year = {2016},
  month = {Apr},
  publisher = {American Physical Society},
  doi = {10.1103/PhysRevLett.116.150502},
  url = {https://link.aps.org/doi/10.1103/PhysRevLett.116.150502}
}

@article{PhysRevLett.117.030401,
  title = {Critical Examination of Incoherent Operations and a Physically Consistent Resource Theory of Quantum Coherence},
  author = {Chitambar, Eric and Gour, Gilad},
  journal = {Phys. Rev. Lett.},
  volume = {117},
  issue = {3},
  pages = {030401},
  numpages = {5},
  year = {2016},
  month = {Jul},
  publisher = {American Physical Society},
  doi = {10.1103/PhysRevLett.117.030401},
  url = {https://link.aps.org/doi/10.1103/PhysRevLett.117.030401}
}

@article{RevModPhys.89.041003,
  title = {Colloquium: Quantum coherence as a resource},
  author = {Streltsov, Alexander and Adesso, Gerardo and Plenio, Martin B.},
  journal = {Rev. Mod. Phys.},
  volume = {89},
  issue = {4},
  pages = {041003},
  numpages = {34},
  year = {2017},
  month = {Oct},
  publisher = {American Physical Society},
  doi = {10.1103/RevModPhys.89.041003},
  url = {https://link.aps.org/doi/10.1103/RevModPhys.89.041003}
}

@article{PhysRevLett.120.230504,
  title = {Experimental Demonstration of Observability and Operability of Robustness of Coherence},
  author = {Zheng, Wenqiang and Ma, Zhihao and Wang, Hengyan and Fei, Shao-Ming and Peng, Xinhua},
  journal = {Phys. Rev. Lett.},
  volume = {120},
  issue = {23},
  pages = {230504},
  numpages = {5},
  year = {2018},
  month = {Jun},
  publisher = {American Physical Society},
  doi = {10.1103/PhysRevLett.120.230504},
  url = {https://link.aps.org/doi/10.1103/PhysRevLett.120.230504}
}

@article{hu2018quantum,
  title={Quantum coherence and geometric quantum discord},
  author={Hu, Ming-Liang and Hu, Xueyuan and Wang, Jieci and Peng, Yi and Zhang, Yu-Ran and Fan, Heng},
  journal={Phys. Rep.},
  volume={762},
  pages={1--100},
  year={2018},
  publisher={Elsevier},
  doi={https://doi.org/10.1016/j.physrep.2018.07.004}
}

@article{wu2021experimental,
  title={Experimental progress on quantum coherence: detection, quantification, and manipulation},
  author={Wu, Kang-Da and Streltsov, Alexander and Regula, Bartosz and Xiang, Guo-Yong and Li, Chuan-Feng and Guo, Guang-Can},
  journal={Adv. Quantum Technol.},
  volume={4},
  number={9},
  pages={2100040},
  year={2021},
  publisher={Wiley Online Library},
  doi={https://doi.org/10.1002/qute.202100040}
}

@article{PhysRevLett.132.180202,
  title = {Arbitrary Amplification of Quantum Coherence in Asymptotic and Catalytic Transformation},
  author = {Shiraishi, Naoto and Takagi, Ryuji},
  journal = {Phys. Rev. Lett.},
  volume = {132},
  issue = {18},
  pages = {180202},
  numpages = {7},
  year = {2024},
  month = {May},
  publisher = {American Physical Society},
  doi = {10.1103/PhysRevLett.132.180202},
  url = {https://link.aps.org/doi/10.1103/PhysRevLett.132.180202}
}

@article{PhysRevA.109.052443,
  title = {Coherence quantifier based on the quantum optimal transport cost},
  author = {Shi, Xian},
  journal = {Phys. Rev. A},
  volume = {109},
  issue = {5},
  pages = {052443},
  numpages = {7},
  year = {2024},
  month = {May},
  publisher = {American Physical Society},
  doi = {10.1103/PhysRevA.109.052443},
  url = {https://link.aps.org/doi/10.1103/PhysRevA.109.052443}
}

@article{PhysRevA.110.042425,
  title = {Quantum speed limit in terms of coherence variations},
  author = {Mai, Zi-yi and Yu, Chang-shui},
  journal = {Phys. Rev. A},
  volume = {110},
  issue = {4},
  pages = {042425},
  numpages = {9},
  year = {2024},
  month = {Oct},
  publisher = {American Physical Society},
  doi = {10.1103/PhysRevA.110.042425},
  url = {https://link.aps.org/doi/10.1103/PhysRevA.110.042425}
}

@article{PhysRevA.54.3824,
  title = {Mixed-state entanglement and quantum error correction},
  author = {Bennett, Charles H. and DiVincenzo, David P. and Smolin, John A. and Wootters, William K.},
  journal = {Phys. Rev. A},
  volume = {54},
  issue = {5},
  pages = {3824--3851},
  numpages = {0},
  year = {1996},
  month = {Nov},
  publisher = {American Physical Society},
  doi = {10.1103/PhysRevA.54.3824},
  url = {https://link.aps.org/doi/10.1103/PhysRevA.54.3824}
}

@article{PhysRevA.100.032324,
  title = {Quantum coherence fraction},
  author = {Yao, Yao and Li, Dong and Sun, C. P.},
  journal = {Phys. Rev. A},
  volume = {100},
  issue = {3},
  pages = {032324},
  numpages = {7},
  year = {2019},
  month = {Sep},
  publisher = {American Physical Society},
  doi = {10.1103/PhysRevA.100.032324},
  url = {https://link.aps.org/doi/10.1103/PhysRevA.100.032324}
}

@article{grondalski2002fully,
  title={The fully entangled fraction as an inclusive measure of entanglement applications},
  author={Grondalski, J and Etlinger, DM and James, DFV},
  journal={Phys. Lett. A},
  volume={300},
  number={6},
  pages={573--580},
  year={2002},
  publisher={Elsevier},
  doi={https://doi.org/10.1016/S0375-9601(02)00884-8}
}

@article{PhysRevA.62.012311,
  title = {Local environment can enhance fidelity of quantum teleportation},
  author = {Badzia\ifmmode \mbox{\c{}}\else \c{}\fi{}g, Piotr and Horodecki, Micha\l{} and Horodecki, Pawe\l{} and Horodecki, Ryszard},
  journal = {Phys. Rev. A},
  volume = {62},
  issue = {1},
  pages = {012311},
  numpages = {7},
  year = {2000},
  month = {Jun},
  publisher = {American Physical Society},
  doi = {10.1103/PhysRevA.62.012311},
  url = {https://link.aps.org/doi/10.1103/PhysRevA.62.012311}
}

@article{PhysRevA.66.012301,
  title = {Optimal teleportation based on bell measurements},
  author = {Albeverio, Sergio and Fei, Shao-Ming and Yang, Wen-Li},
  journal = {Phys. Rev. A},
  volume = {66},
  issue = {1},
  pages = {012301},
  numpages = {4},
  year = {2002},
  month = {Jul},
  publisher = {American Physical Society},
  doi = {10.1103/PhysRevA.66.012301},
  url = {https://link.aps.org/doi/10.1103/PhysRevA.66.012301}
}

@article{PhysRevLett.127.080502,
  title = {Catalytic Quantum Teleportation},
  author = {Lipka-Bartosik, Patryk and Skrzypczyk, Paul},
  journal = {Phys. Rev. Lett.},
  volume = {127},
  issue = {8},
  pages = {080502},
  numpages = {7},
  year = {2021},
  month = {Aug},
  publisher = {American Physical Society},
  doi = {10.1103/PhysRevLett.127.080502},
  url = {https://link.aps.org/doi/10.1103/PhysRevLett.127.080502}
}

@article{PhysRevA.94.062120,
  title = {Quantum steerability: Characterization, quantification, superactivation, and unbounded amplification},
  author = {Hsieh, Chung-Yun and Liang, Yeong-Cherng and Lee, Ray-Kuang},
  journal = {Phys. Rev. A},
  volume = {94},
  issue = {6},
  pages = {062120},
  numpages = {13},
  year = {2016},
  month = {Dec},
  publisher = {American Physical Society},
  doi = {10.1103/PhysRevA.94.062120},
  url = {https://link.aps.org/doi/10.1103/PhysRevA.94.062120}
}

@article{PhysRevA.94.062123,
  title = {Superactivation of quantum steering},
  author = {Quintino, Marco T\'ulio and Brunner, Nicolas and Huber, Marcus},
  journal = {Phys. Rev. A},
  volume = {94},
  issue = {6},
  pages = {062123},
  numpages = {7},
  year = {2016},
  month = {Dec},
  publisher = {American Physical Society},
  doi = {10.1103/PhysRevA.94.062123},
  url = {https://link.aps.org/doi/10.1103/PhysRevA.94.062123}
}

@article{PhysRevA.96.012107,
  title = {Work extraction and fully entangled fraction},
  author = {Hsieh, Chung-Yun and Lee, Ray-Kuang},
  journal = {Phys. Rev. A},
  volume = {96},
  issue = {1},
  pages = {012107},
  numpages = {8},
  year = {2017},
  month = {Jul},
  publisher = {American Physical Society},
  doi = {10.1103/PhysRevA.96.012107},
  url = {https://link.aps.org/doi/10.1103/PhysRevA.96.012107}
}

@article{bai2015maximallycoherentstates,
author = {Bai, Zhaofang and Du, Shuanping},
title = {Maximally coherent states},
year = {2015},
issue_date = {November 2015},
publisher = {Rinton Press, Incorporated},
address = {Paramus, NJ},
volume = {15},
number = {15–16},
issn = {1533-7146},
journal = {Quantum Info. Comput.},
month = nov,
pages = {1355–1364},
numpages = {10},
doi={https://doi.org/10.48550/arXiv.1503.07103}
}

@article{PhysRevA.93.032326,
  title = {Maximally coherent states and coherence-preserving operations},
  author = {Peng, Yi and Jiang, Yong and Fan, Heng},
  journal = {Phys. Rev. A},
  volume = {93},
  issue = {3},
  pages = {032326},
  numpages = {6},
  year = {2016},
  month = {Mar},
  publisher = {American Physical Society},
  doi = {10.1103/PhysRevA.93.032326},
  url = {https://link.aps.org/doi/10.1103/PhysRevA.93.032326}
}

@article{karmakar2019coherence,
  title={Coherence fraction},
  author={Karmakar, Sumana and Sen, Ajoy and Chattopadhyay, Indrani and Bhar, Amit and Sarkar, Debasis},
  journal={Quantum Inf. Process.},
  volume={18},
  number={9},
  pages={275},
  year={2019},
  publisher={Springer},
  doi={https://doi.org/10.1007/s11128-019-2391-6}
}

@article{PhysRevA.92.022124,
  title = {Intrinsic randomness as a measure of quantum coherence},
  author = {Yuan, Xiao and Zhou, Hongyi and Cao, Zhu and Ma, Xiongfeng},
  journal = {Phys. Rev. A},
  volume = {92},
  issue = {2},
  pages = {022124},
  numpages = {8},
  year = {2015},
  month = {Aug},
  publisher = {American Physical Society},
  doi = {10.1103/PhysRevA.92.022124},
  url = {https://link.aps.org/doi/10.1103/PhysRevA.92.022124}
}

@article{jozsa1998quantum,
  title={Quantum algorithms and the Fourier transform},
  author={Jozsa, Richard},
  journal={Phil. Trans. R. Soc. A.},
  volume={454},
  number={1969},
  pages={323--337},
  year={1998},
  publisher={The Royal Society},
  doi = {https://doi.org/10.1098/rspa.1998.0163},
  url = {https://doi.org/10.1098/rspa.1998.0163}
}

@article{ekert1998quantum,
  title={Quantum algorithms: entanglement--enhanced information processing},
  author={Ekert, Artur and Jozsa, Richard},
  journal={Phil. Trans. R. Soc. A.},
  volume={356},
  number={1743},
  pages={1769--1782},
  year={1998},
  publisher={The Royal Society},
  doi = {https://doi.org/10.1098/rsta.1998.0248},
  url = {https://doi.org/10.1098/rsta.1998.0248}
}

@article{PhysRevA.65.062312,
  title = {Entanglement monotone derived from Grover's algorithm},
  author = {Biham, Ofer and Nielsen, Michael A. and Osborne, Tobias J.},
  journal = {Phys. Rev. A},
  volume = {65},
  issue = {6},
  pages = {062312},
  numpages = {7},
  year = {2002},
  month = {Jun},
  publisher = {American Physical Society},
  doi = {10.1103/PhysRevA.65.062312},
  url = {https://link.aps.org/doi/10.1103/PhysRevA.65.062312}
}

@article{PhysRevA.93.012111,
  title = {Coherence as a resource in decision problems: The Deutsch-Jozsa algorithm and a variation},
  author = {Hillery, Mark},
  journal = {Phys. Rev. A},
  volume = {93},
  issue = {1},
  pages = {012111},
  numpages = {6},
  year = {2016},
  month = {Jan},
  publisher = {American Physical Society},
  doi = {10.1103/PhysRevA.93.012111},
  url = {https://link.aps.org/doi/10.1103/PhysRevA.93.012111}
}

@article{PhysRevA.95.032307,
  title = {Coherence depletion in the Grover quantum search algorithm},
  author = {Shi, Hai-Long and Liu, Si-Yuan and Wang, Xiao-Hui and Yang, Wen-Li and Yang, Zhan-Ying and Fan, Heng},
  journal = {Phys. Rev. A},
  volume = {95},
  issue = {3},
  pages = {032307},
  numpages = {8},
  year = {2017},
  month = {Mar},
  publisher = {American Physical Society},
  doi = {10.1103/PhysRevA.95.032307},
  url = {https://link.aps.org/doi/10.1103/PhysRevA.95.032307}
}

@article{PhysRevA.100.012349,
  title = {Operator coherence dynamics in Grover's quantum search algorithm},
  author = {Pan, Minghua and Qiu, Daowen},
  journal = {Phys. Rev. A},
  volume = {100},
  issue = {1},
  pages = {012349},
  numpages = {10},
  year = {2019},
  month = {Jul},
  publisher = {American Physical Society},
  doi = {10.1103/PhysRevA.100.012349},
  url = {https://link.aps.org/doi/10.1103/PhysRevA.100.012349}
}

@article{pan2019entangling,
  title={Entangling and disentangling in Grover's search algorithm},
  author={Pan, Minghua and Qiu, Daowen and Mateus, Paulo and Gruska, Jozef},
  journal={Theor. Comput. Sci.},
  volume={773},
  pages={138--152},
  year={2019},
  publisher={Elsevier},
  doi={https://doi.org/10.1016/j.tcs.2018.10.001}
}

@article{pan2022complementarity,
  title={Complementarity between success probability and coherence in Grover search algorithm},
  author={Pan, Minghua and Situ, Haozhen and Zheng, Shenggen},
  journal={EPL},
  volume={138},
  number={4},
  pages={48002},
  year={2022},
  publisher={IOP Publishing},
  doi={10.1209/0295-5075/ac7165}
}

@article{PhysRevA.106.062429,
  title = {Entanglement and coherence in the Bernstein-Vazirani algorithm},
  author = {Naseri, Moein and Kondra, Tulja Varun and Goswami, Suchetana and Fellous-Asiani, Marco and Streltsov, Alexander},
  journal = {Phys. Rev. A},
  volume = {106},
  issue = {6},
  pages = {062429},
  numpages = {13},
  year = {2022},
  month = {Dec},
  publisher = {American Physical Society},
  doi = {10.1103/PhysRevA.106.062429},
  url = {https://link.aps.org/doi/10.1103/PhysRevA.106.062429}
}

@article{PhysRevLett.129.120501,
  title = {Coherence as a Resource for Shor's Algorithm},
  author = {Ahnefeld, Felix and Theurer, Thomas and Egloff, Dario and Matera, Juan Mauricio and Plenio, Martin B.},
  journal = {Phys. Rev. Lett.},
  volume = {129},
  issue = {12},
  pages = {120501},
  numpages = {7},
  year = {2022},
  month = {Sep},
  publisher = {American Physical Society},
  doi = {10.1103/PhysRevLett.129.120501},
  url = {https://link.aps.org/doi/10.1103/PhysRevLett.129.120501}
}

@inproceedings{bernstein1993quantum,
  title={Quantum complexity theory},
  author={Bernstein, Ethan and Vazirani, Umesh},
  booktitle={Proceedings of the twenty-fifth annual ACM symposium on Theory of computing},
  pages={11--20},
  year={1993}
}

@article{bernstein1997quantum,
  title={Quantum Complexity Theory},
  author={Bernstein, Ethan and Vazirani, Umesh},
  journal={SIAM J. Comput.},
  volume={26},
  number={5},
  pages={1411--1473},
  year={1997},
  publisher={SIAM},
  doi={https://doi.org/10.1137/S0097539796300921}
}

@article{bravyi2018quantum,
  title={Quantum advantage with shallow circuits},
  author={Bravyi, Sergey and Gosset, David and K{\"o}nig, Robert},
  journal={Science},
  volume={362},
  number={6412},
  pages={308--311},
  year={2018},
  publisher={American Association for the Advancement of Science},
  doi={10.1126/science.aar3106}
}

@article{deutsch1985quantum,
  title={Quantum theory, the Church--Turing principle and the universal quantum computer},
  author={Deutsch, David},
  journal={Proc. R. Soc. Lond. A},
  volume={400},
  number={1818},
  pages={97--117},
  year={1985},
  publisher={The Royal Society London},
  doi={https://doi.org/10.1098/rspa.1985.0070}
}

@article{deutsch1992rapid,
  title={Rapid solution of problems by quantum computation},
  author={Deutsch, David and Jozsa, Richard},
  journal={Proc. R. Soc. Lond. A},
  volume={439},
  number={1907},
  pages={553--558},
  year={1992},
  publisher={The Royal Society London},
  doi={https://doi.org/10.1098/rspa.1992.0167}
}

@article{xie2018quantum,
  title={Quantum algorithms on Walsh transform and Hamming distance for Boolean functions},
  author={Xie, Zhengwei and Qiu, Daowen and Cai, Guangya},
  journal={Quantum Inf. Process.},
  volume={17},
  pages={1--17},
  year={2018},
  publisher={Springer},
  doi={https://doi.org/10.1007/s11128-018-1885-y}
}

@article{zhou2023distributed,
  title={Distributed Bernstein--Vazirani algorithm},
  author={Zhou, Xu and Qiu, Daowen and Luo, Le},
  journal={Phys. A: Stat. Mech. its Appl.},
  volume={629},
  pages={129209},
  year={2023},
  publisher={Elsevier},
  doi={https://doi.org/10.1016/j.physa.2023.129209}
}

@article{PhysRevLett.130.210602,
  title = {Demonstration of Algorithmic Quantum Speedup},
  author = {Pokharel, Bibek and Lidar, Daniel A.},
  journal = {Phys. Rev. Lett.},
  volume = {130},
  issue = {21},
  pages = {210602},
  numpages = {6},
  year = {2023},
  month = {May},
  publisher = {American Physical Society},
  doi = {10.1103/PhysRevLett.130.210602},
  url = {https://link.aps.org/doi/10.1103/PhysRevLett.130.210602}
}

@article{uhlmann1976The,
    title = {The “transition probability” in the state space of a $*$-algebra},
    author = {A. Uhlmann},
    journal = {Rep. Math. Phys.},
    volume = {9},
    number = {2},
    pages = {273-279},
    year = {1976},
    issn = {0034-4877},
    doi = {https://doi.org/10.1016/0034-4877(76)90060-4},
    url = {https://www.sciencedirect.com/science/article/pii/0034487776900604}
}

@article{Jozsa1994Fidelity,
author = {Richard Jozsa},
title = {Fidelity for Mixed Quantum States},
journal = {J. Mod. Opt},
volume = {41},
number = {12},
pages = {2315--2323},
year = {1994},
publisher = {Taylor \& Francis},
doi = {10.1080/09500349414552171},
URL = {https://doi.org/10.1080/09500349414552171}
}

@article{Liang2019Quantum,
doi = {10.1088/1361-6633/ab1ca4},
year = {2019},
month = {jun},
publisher = {IOP Publishing},
volume = {82},
number = {7},
pages = {076001},
author = {Yeong-Cherng Liang and Yu-Hao Yeh and Paulo E M F Mendonça and Run Yan Teh and Margaret D Reid and Peter D Drummond},
title = {Quantum fidelity measures for mixed states},
journal = {Rep. Prog. Phys.},
url = {https://dx.doi.org/10.1088/1361-6633/ab1ca4}
}

@article{PhysRevA.110.062429,
  title = {Coherence fraction in Grover's search algorithm},
  author = {Zhou, Si-Qi and Jin, Hai and Liang, Jin-Min and Fei, Shao-Ming and Xiao, Yunlong and Ma, Zhihao},
  journal = {Phys. Rev. A},
  volume = {110},
  issue = {6},
  pages = {062429},
  numpages = {8},
  year = {2024},
  month = {Dec},
  publisher = {American Physical Society},
  doi = {10.1103/PhysRevA.110.062429},
  url = {https://link.aps.org/doi/10.1103/PhysRevA.110.062429}
}

@article{zhang2024quantification,
  title={Quantification of entanglement and coherence with purity detection},
  author={Zhang, Ting and Smith, Graeme and Smolin, John A and Liu, Lu and Peng, Xu-Jie and Zhao, Qi and Girolami, Davide and Ma, Xiongfeng and Yuan, Xiao and Lu, He},
  journal={npj Quantum Inform.},
  volume={10},
  number={1},
  pages={60},
  year={2024},
  publisher={Nature Publishing Group UK London},
  doi={https://doi.org/10.1038/s41534-024-00857-2}
}

@article{duan2001long,
  title={Long-distance quantum communication with atomic ensembles and linear optics},
  author={Duan, L-M and Lukin, Mikhail D and Cirac, J Ignacio and Zoller, Peter},
  journal={Nature},
  volume={414},
  number={6862},
  pages={413--418},
  year={2001},
  publisher={Nature Publishing Group UK London},
  doi={https://doi.org/10.1038/35106500}
}

@INPROCEEDINGS{365700,
  author={Shor, P.W.},
  booktitle={Proceedings 35th Annual Symposium on Foundations of Computer Science}, 
  title={Algorithms for quantum computation: discrete logarithms and factoring}, 
  year={1994},
  volume={},
  number={},
  pages={124-134},
  doi={10.1109/SFCS.1994.365700}}

\end{document}